\documentclass[jkps,preprint,showpacs,preprintnumbers,amsmath,amssymb,showkeys]{revtex4}
\usepackage{graphicx}
\usepackage{dcolumn}
\usepackage{bm}
\usepackage{ulem}

\begin{document}
\title{ Gap states and edge properties of   rectangular  graphene quantum dot in staggered potential}
\author{Y. H. Jeong and S. -R. Eric Yang
\footnote{corresponding author, eyang812@gmail.com}  }
\affiliation{  Department of Physics, Korea  University, Seoul,
Korea\\}

\begin{abstract}
We investigate edge properties of a gapful rectangular graphene quantum dot   in a staggered potential.  In such a system
gap states with discrete and  closely spaced energy levels exist that are spatially  located on  the left or right zigzag edge. We find that, although the bulk states outside the energy gap are nearly unaffected, spin degeneracy of
each gap state is lifted by the staggered potential.
We have computed  the occupation numbers of  spin-up and -down gap states at various values of the strength of the staggered potential.
The   electronic and magnetic properties of the zigzag edges depend sensitively on these numbers.  We discuss the possibility of applying this system as a single electron spintronic device.
\end{abstract}


\maketitle

\section{Introduction}

Graphene exhibits interesting fundamental physics\cite{Neto}, such as quantum Hall effect\cite{PKim}, and  Berry\cite{Ando} and Zak phases. Graphene nanostructures are also important building blocks for device applications\cite{Neto,Loss,Shin,Sch}.  In this paper we focus on  a nanostructure of
a rectangular graphene quantum dot (RGQD) with   two   armchair edges and two  zigzag edges\cite{Tang,Kim}.   It has several interesting properties.  For certain values of the length of the zigzag edges an excitation gap exists that are filled with topological gap states\cite{JeongJNN}.  These    gap states  are localized on the zigzag edges\cite{Fuj,Son,Jeong}, and their number grows with the length of the zigzag edges $L_{zig}$\cite{Jeong0}.   Half of these gap states are localized on the left edge and the other half on the right edge.
These gap states
are  responsible for antiferromagnetism between opposite zigzag edges\cite{Tang}.   In the presence of a weak magnetic field these gap states are no longer localized entirely on the zigzag edges, and can display unexpected  patterns in their probability densities\cite{Kim}.

\begin{figure}[!hbpt]
\begin{center}
\includegraphics[width=0.35\textwidth]{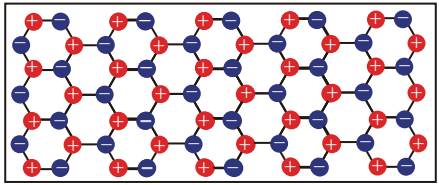}
\caption{ RGQR has  two armchair edges and two zigzag edges.  The lengths of  the armchair and zigzag edges are, respectively, $L_{arm}$ and $L_{zig}$.   The staggered potential energy is  $\epsilon_i=\Delta/2$ on sublattice A (red circles)  and $\epsilon_i=-\Delta/2$ on sublattice B (blue circles). Note that the opposite zigzag edges experience different  potential energies. }
\label{stag}
\end{center}
\end{figure}

In this paper we investigate the effect of a staggered potential of a substrate\cite{stagpot1,stagpot2,stagpot3} on the electronic properties of a RGQD with gap states.     It has  a profound effect on electronic states localized on the zigzag edges. One would expect that, since the opposite zigzag edges experience different electric potentials (see Fig.\ref{stag}),  charge imbalance  occurs between   the left and right zigzag edges, which will in turn affect  edge  antiferromagnetism.  The effect of a staggered potential on the edge states is  analogues to the effect    of  a uniform  electric field, but unlike  it, a staggered potential does not affect extended states significantly due to its alternating sign, as shown in Fig.\ref{stag}. Its effect
has been investigated recently in a periodic zigzag graphene ribbon (PZGR), which has  a band crossing at the Fermi energy $E_F=0$ when  it is   half-filled.  Depending on the strength of electron-electron interactions, interesting band structures  have been found, such as an antiferromagnetic  insulating band  and antiferromagnetic half metallic band with a non-trivial spin structure\cite{Sor}.

In this paper we will explore electronic and magnetic properties of RGQDs with a sizable  excitation gap so that gap states are isolated from  lower and higher energy quasi-continuum bulk states.
These gap states have discrete and  closely spaced energy levels, and they   are spatially  located near  the left or right zigzag edge.  We find that spin degeneracy of each gap state is broken by the staggered potential.
The   electronic and magnetic properties of the zigzag edges depend sensitively on the occupation numbers of spin-up and -down gap states.
Zigzag edges with different charge imbalance and total spin value are possible for different values of the strength of the staggered potential.    For a certain range of the strength of staggered potential $\Delta$ antiferromagnetically coupled zigzag edges  exist, but outside this range the electrons in the gap states are all localized only on one zigzag edge. The physics behind the edge magnetism   is the competition between lowering  of the total  staggered potential energy  and the energy cost  due to the repulsive  interactions  when  electrons move to the edge with the  lower
staggered potential energy.   In contrast to a uniform electric field,  a staggered potential does not change significantly  the extended states outside the gap.

This paper is organized as follows.  In See. II  we define our model.  Using it we compute gap states in Sec.III.   The charge and magnetic configurations of the edges are computed in Sec.IV.  Summary and discussion are given in Sec.V.

\section{  Model }

Since translational symmetry is broken we write the Hamiltonian  in the site representation to compute the groundstate of a RGQD.  We adopt a tight-binding model of a RGQD at half-filling with the on-site repulsion $U$ and solve it using the Hartree-Fock (HF)  approximation (this approach is widely used and its results are consistent with those of DFT\cite{Fuj,Pis,Yang} ).  The HF Hamiltonian is
\begin{eqnarray}
&&H=-t\sum_{ij\sigma} c_{i\sigma}^{\dag}c_{j\sigma}+\sum_{i\sigma} \epsilon_ic_{i\sigma}^{\dag}c_{i\sigma}+
U\sum_{i\sigma}  (n_{i\uparrow} \langle n_{i\downarrow}\rangle+\langle n_{i\uparrow}\rangle n_{i\downarrow}
-\langle n_{i\uparrow}\rangle \langle n_{i\downarrow}\rangle ),
\label{Zak}
\end{eqnarray}
where $\epsilon_i=\pm \Delta/2$, $c_{i\sigma}^{\dag}$ and $n_{i\sigma}$ are the value of the staggered potential at site $i$,  electron creation and  occupation operators at site $i$ with spin $\sigma$.   In the hopping term the summation is over the nearest neighbor sites (the hopping parameter $t\sim 3$ eV).  When
the length of the zigzag  edges is
$ L_{zig}=(3M + 1)a_0$ or $3Ma_0$ an excitation gap exist\cite{exptarm,Brey,Lee,Yang}  ($M$ is an integer and $a_0$ is graphene unit cell length).  When graphene is epitaxially grown on SiC substrate there is an additional contribution $\Gamma \sim 0.26$ eV to the excitation gap due to the staggered potential \cite{stagpot1}.  A ribbon with $L_{zig} = (3M + 2)a_0$ does not have a gap in the absence of electron-electron interactions and will not be considered   here.
We consider  RGQDs  with  width in the range $\frac{L_{zig}}{L_{arm}}< 1$  (when  $\frac{L_{zig}}{L_{arm}}\gg1$ a zigzag ribbon is realized; in this case gap state energies are not well isolated from each other and from the quasi-continuous conduction and valence band states).  In general the nature of the gap/edge states state depends on the interplay between several parameters $\Delta$, $U$, $t$, $L_{zig}$,  and $L_{arm}$.

\section{Energy levels of gap states   }

It is useful to define  the  number of  occupied {\it gap states} of spin $\sigma$ located on the left (right) edge $N_{L (R),\sigma}$.
In addition, it is helpful to compute the following  sums of the   occupation numbers for each edge:  $N_{L}=N_{L,\uparrow}+N_{L,\downarrow}$ for the left edge and
$N_{R}=N_{R,\uparrow}+N_{R,\downarrow}$ for the right edge.  The    differences of the  spin occupation  numbers for each edge are defined as
$M_{L}=N_{L,\uparrow}-N_{L,\downarrow}$ for the left edge and
$M_{R}=N_{R,\uparrow}-N_{R,\downarrow}$ for the right edge.

We compute the  energy levels of gap states   of a RGQD with
$L_{zig}=22.41${\AA}   and $L_{arm}=215.84${\AA}.
For this RGQD there are   $12$ gap/end states (counting spins).  There are $9$ end sites on each zigzag edge.
On each edge,  the occupied   quasi-continuum states  contribute $6$ electrons and the end states
contribute $3$ additional electrons, and    there are $3+6=9$ electrons in total (note that the system is half-filled and each edge has $9$ sites).

\begin{figure}[!hbpt]
\begin{center}
\includegraphics[width=0.2\textwidth]{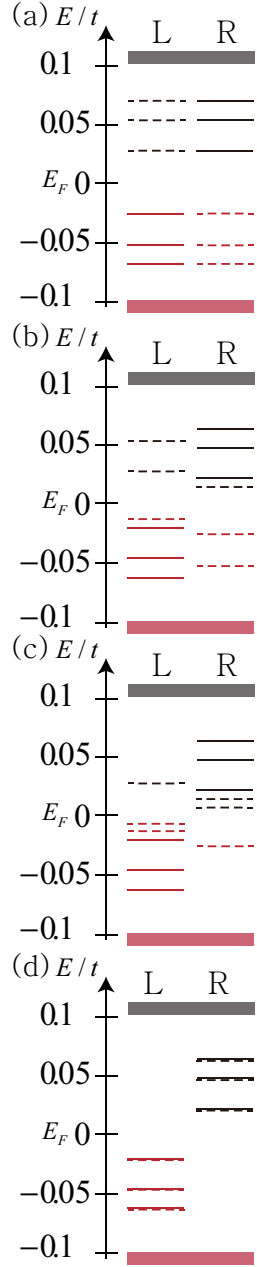}
\caption{ Schematic energy levels of $12$ spin-up and -down gap states are shown at various values of $\Delta/t=0, 0.026, 0.0267, 0.08$ (increasing from (a) to (d)).    Solid (dashed) lines represent spin-up (down) gap states. They are all localized on either left (L) or right (R)  zigzag edge.   On the left edge occupied spin-down gap states increases from $0,1,2$ to $3$ while on the right edge occupied spin-up gap states decreases from $3,2,1$ to $0$.  Note that the Fermi energy is $E_F=0$.
Shaded areas represent quasi-continuum bulk states. The change of the energy position of these  states due to the staggered potential is negligible.  The energy spacing between the gap states is  $\sim 0.01t$. Parameters are $L_x=22.41$\AA, $L_y=215.84$, and $U=0.5t$.  Even in the absence of a staggered potential there will be an energy
gap and gap states.   Outside this gap one has quasi-continuous
conduction and valence bands.  }
\label{Thick}
\end{center}
\end{figure}

\begin{figure}[!hbpt]
\begin{center}
\includegraphics[width=0.2\textwidth]{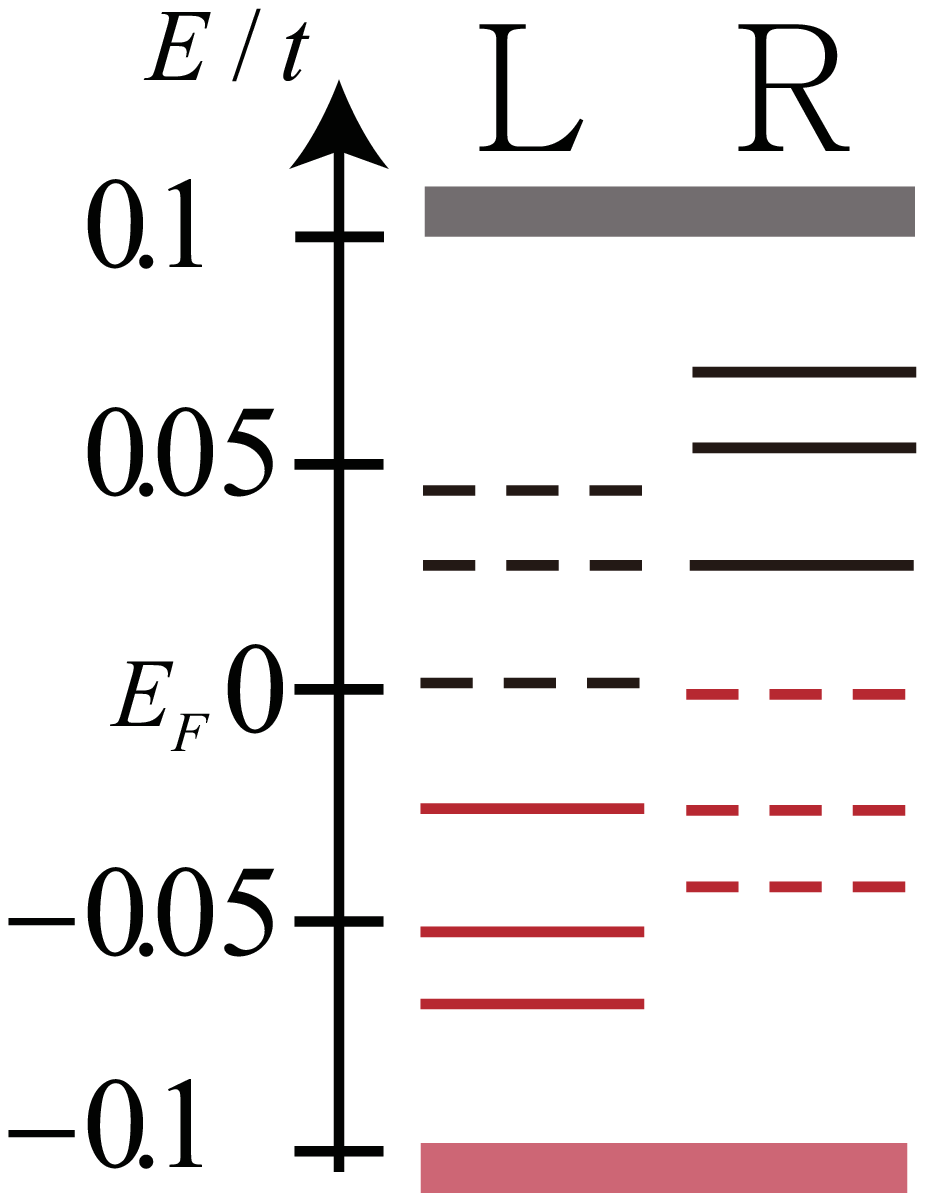}
\caption{ Schematic energy levels of $12$ spin-up and -down gap states are shown at the transition point values  $\Delta=0.226t$. Note that two spin-down gap states are degenerate at the Fermi energy.  Parameters are same as those given in Fig.\ref{Thick}.  }
\label{transition}
\end{center}
\end{figure}

Figure \ref{Thick}(a)  displays the energy spectrum of gap states
at $\Delta=0$.  The numbers of the occupied gap states on the
left edge  are $(N_{L,\uparrow},N_{L,\downarrow})=(3,0)$  and on the right edge   $(N_{R,\uparrow},N_{R,\downarrow})=(0,3)$.    The system is
antiferromagnetic with  $M_L=3$   and $M_R=-3$ (this antiferromagnetic state is almost degenerate with a ferromagnetic state\cite{Tang}).  There  is   no charge imbalance since  $N_L=3$  and  $N_R=3$.   One spin-down electron transfers  discontinuously just after $\Delta=0.026t $.  At the transition point
two spin-down gap states become degenerate at the Fermi energy.   One of them is the highest energy
occupied spin-down gap state, that is localized on the left zigzag edge, see Fig.\ref{transition}.     The other is
the lowest energy unoccupied spin-down gap state that is localized on the right zigzag edge. As we can see Figs.\ref{Thick}  and \ref{transition} the average energy spacing between the gap states located on the left and right edges becomes smaller
as a transition point is approached.

Figure \ref{Thick}(b)  displays the schematic energy spectrum of gap states after  this transition.   Note that the energy spacing between the gap states {\it does not vary uniformly} with increasing $\Delta$.
The gap occupation numbers are  $(N_{L,\uparrow},N_{L,\downarrow})=(3,1)$ and $(N_{R,\uparrow},N_{R,\downarrow})=(0,2)$.   The charge imbalance increases: $N_L=4$  and  $N_R=2$.  But   antiferromagnetism is weakened by this process since the magnitude of total spin on each edge is reduced:  $M_L=2$   and $M_R=-2$.

There are other two discontinuous changes near  $\Delta=0.069t$ and   $0.073t$.  Figures \ref{Thick}(c)  and (d) display the energy spectra of gap states after these transitions.
Let us discuss the third  electron transfer  near $\Delta=0.073t$.   Just before the transition the numbers of occupied gap states on the
left edge  are $(N_{L,\uparrow},N_{L,\downarrow})=(3,2)$  and on the right edge   $(N_{R,\uparrow},N_{R,\downarrow})=(0,1)$.
Since $M_L=1$  and    $M_R=-1$  the zigzag edges  display antiferromagnetism.   There is also a  charge imbalance:  $N_L=5$  and  $N_R=1$.
After  the  transition,
one spin-down edge electron is transferred  from  the right end to the left end:   the numbers of occupied gap states on the
left edge  are $(N_{L,\uparrow},N_{L,\downarrow})=(3,3)$  and on the right edge   $(N_{R,\uparrow},N_{R,\downarrow})=(0,0)$.
The system is  now paramagnetic:  $M_L=0$  with the   maximum charge imbalance  $N_L=6$  and  $N_R=0$.  We will call this value the critical value of the staggered potential $\Delta_c=0.073t$ for $U=0.5t$.

\begin{figure}[!hbpt]
\begin{center}
\includegraphics[width=0.4\textwidth]{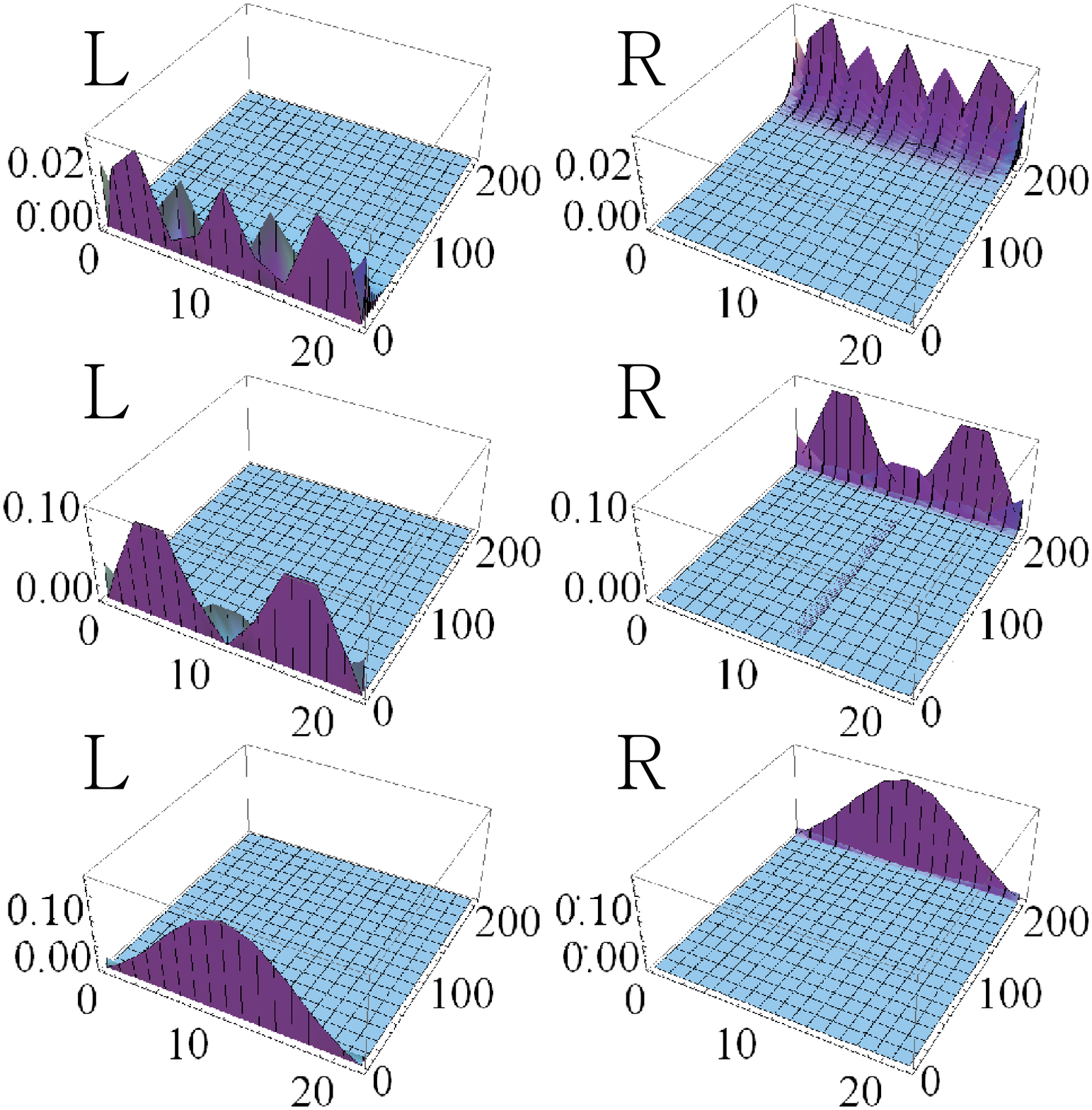}
\caption{ Probability densities of occupied spin-up (left panel)   and -down (right panel) gap/edge states are shown (unit of length  is \AA).  They are all localized on either left (L) or right (R)  zigzag edge.   Other parameters are $L_x=22.41$\AA, $L_y=215.84$, $\Delta = 0.026t$ and $U=0.5t$.  }
\label{Thick1}
\end{center}
\end{figure}

Probability densities of $3$ occupied spin-up gap states localized  on the left edge and $3$ occupied spin-down gap states localized on the right edge
are also shown in Fig.\ref{Thick1}.

\section{Electronic and magnetic properties of zigzag edges}

We have determined the occupation numbers of  the gap states at various values of $\Delta$.   Using these numbers  we will now calculate electronic and magnetic properties of the zigzag edges.
For each spin we compute the average end occupation number  per A-site of   the right zigzag edge
$n_{A,\sigma}$.    Note that, in addition to occupied gap states, there are also occupied quasi-continuum bulk states that have to be included in computing this quantity.  At a B-site on the opposite end  the occupation number $n_{B,\sigma}$   is defined similarly.
The average occupation numbers  per site of the left and right edges are
$n_L=n_{A,\uparrow}+n_{A,\downarrow}$  and
$n_R=n_{B,\uparrow}+n_{B,\downarrow}$.   Similarly
the average magnetizations per site are
$m_L=n_{A,\uparrow}-n_{A,\downarrow}$  and
$m_R=n_{B,\uparrow}-n_{B,\downarrow}$.

\begin{figure}[!hbpt]
\begin{center}
\includegraphics[width=0.45\textwidth]{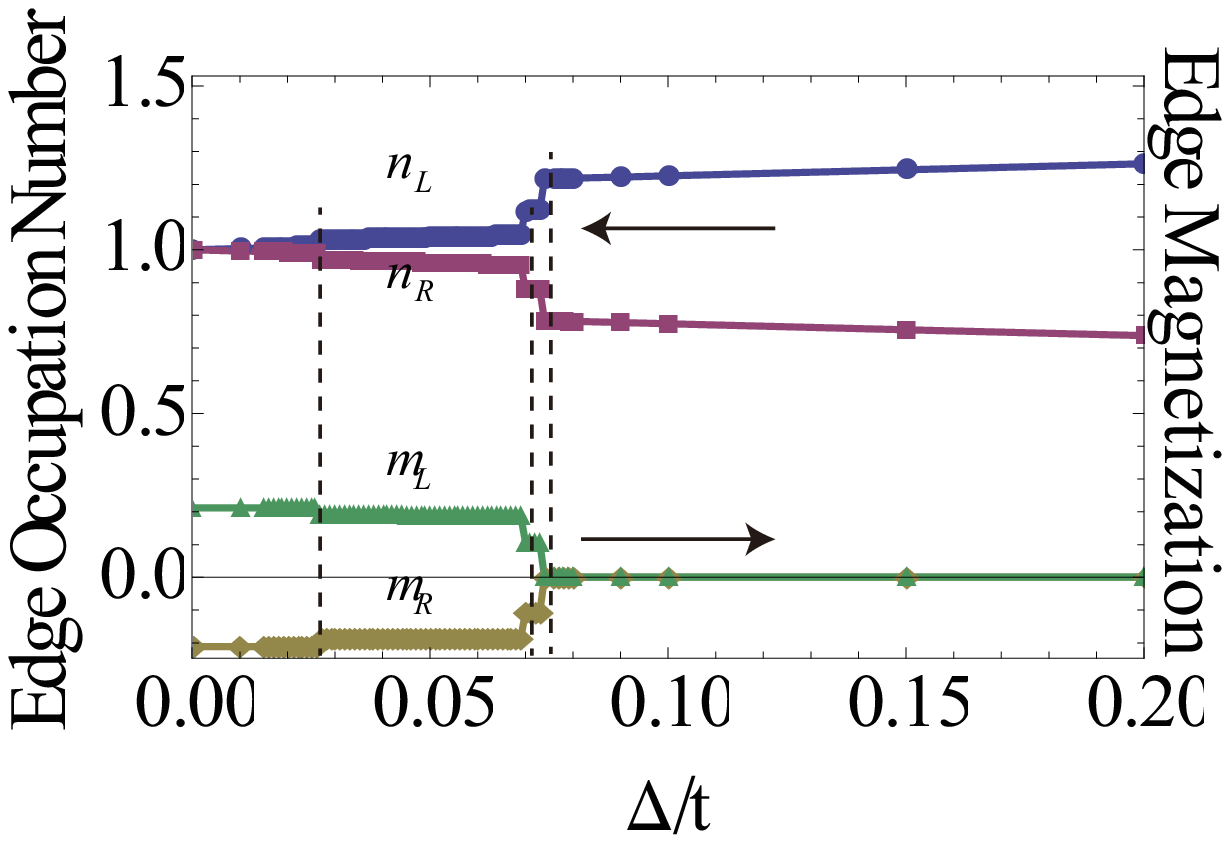}
\caption{ Edge electron occupations  per right end site (filled squares) and   per left end site (filled circles) of  a RGQD  are  plotted as a function of  $\Delta$.  Also plotted are the the magnetizations per left end site (diamonds) and per right end site  (triangles). A vertical line indicates a value of $\Delta$ where a discontinuous  edge reconstruction takes place.  On-site potential is $U=0.5t$.
Dot length is  $L_{arm}=215.84$ \AA    and width is  $L_{zig}=22.41$\AA .
 }
\label{OccDel}
\end{center}
\end{figure}

\begin{figure}[!hbpt]
\begin{center}
\includegraphics[width=0.45\textwidth]{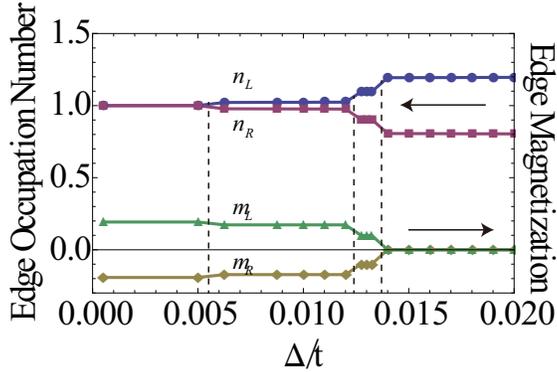}
\caption{ Same as in Fig.\ref{OccDel} but for  $U=0.1t$.
 }
\label{OccDel1}
\end{center}
\end{figure}

Figure \ref{OccDel} displays $n_L$,  $n_R$,   $m_L$, and  $m_R$ for width $22.41$\AA, which we used in the last section.   At   $\Delta=0$,   six    gap states  are occupied: $3$ on the left edge and the other $3$ on the right edge.  In addition, on each edge $6$ other electrons  are in the occupied quasi-continuum bulk states.  So in total    $9$ electrons occupy each edge.
Three discontinuous transitions are present in the variation of  $n_L$,  $n_R$,   $m_L$, and  $m_R$.
At each discontinuous transition a spin-down edge electron transfers from the right to left edge.   After each transition the average occupation per site is approximately  $10/9$, $11/9$, and    $12/9$.   The modification of   the occupied quasi-continuum states in response
to the staggered potential also contribute to the edge occupation numbers, but only slightly:  they contribute to  the slowly changing part of $n_L$,  $n_R$,   $m_L$, and  $m_R$.   Figure \ref{OccDel1} displays $n_L$,  $n_R$,   $m_L$, and  $m_R$ for a smaller value $U=0.1t$.   Note that the critical value
$\Delta_c= 0.014 t$ is smaller than that of $U=0.5t$.  As in Fig.\ref{OccDel} three discontinuous electron transfers are present.

\begin{figure}[!hbpt]
\begin{center}
\includegraphics[width=0.3\textwidth]{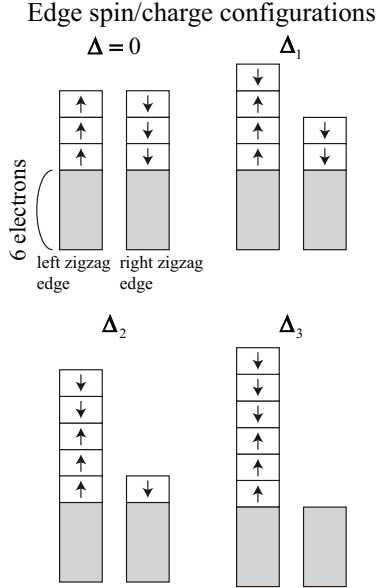}
\caption{Schematic display of edge spin and charge configurations. Shaded areas represent electrons in the quasi-continuum states and boxes represent gap/edge states.   One edge electron moves from the right to left edges at three different values of $\Delta=\Delta_1,\Delta_2$,   and $\Delta_3$.   Shaded areas represent electrons in   the quasi-continuum states.}
\label{conf}
\end{center}
\end{figure}

\section{Summary and discussion}

We have computed the  closely spaced energy levels of gap states of a RGQD, which are located on  the zigzag edges.  We find that, although the energy position of the bulk states outside the energy gap is nearly unaffected, spin degeneracy of
each gap state is broken by the staggered potential.
We have computed  the occupation numbers of  spin-up and -down gap states  at various values of the strength of the staggered potential.  Using them we have determined electronic and magnetic properties of zigzag edges.  They depend sensitively on the value of the staggered potential.
The physical origin
of this effect is that an electron on  the edge with the   high potential energy  tend to move to the low potential energy edge, but the electrons on the low potential energy region tend to repel each other.
Figure \ref{conf} recaptures possible edge electronic and magnetic configurations.    At $\Delta=0$ there is no charge imbalance between the edges.  Charge imbalance increases suddenly at $\Delta_1$,    $\Delta_2$  and $\Delta_3$.  The magnitude of the total spin number on the left edge is $M_L=3$,  $2$,  and $1$ at $\Delta=0, \Delta_1$,  and  $\Delta_2$, respectively.  On the right edge spin values $M_R$ are opposite of these, i.e., they are coupled antiferromagnetically with the spins on the left edge.  At each transition one spin-down edge electron moves to the left zigzag edge.
After the critical value $\Delta_c=\Delta_3$ antiferromagnetism disappears and $M_L=M_R=0$.     The number of transitions depend on the number of gap states, which is determined by the length of the zigzag edges.

We note several special features of the energy spectrum of the gap states.  Unlike a uniform electric field a  staggered potential does not affect significantly the quasi-continuum bulk states outside of the gap.  Moreover,  the energy spacing between the gap states does not vary uniformly with increasing $\Delta$.  In a staggered potential the average energy spacing between the gap states located on the left and right edges can become smaller,  see Figs.\ref{Thick}  and \ref{transition}.   We have also investigated the energy spectrum at several other smaller values of $U/t$.   We find  at      $U/t=0.1$ similar  results  to those of $U=0.5t$, but  the value of the critical strength  $\Delta_c\sim 0.014t$ is smaller than $\Delta_c\sim 0.07t$.     The  properties of the gap states may be probed using scanning tunneling microscopy  measurement of   the differential conductance\cite{Andrei}.

We conclude by mentioning a possible   application of a  RGQD in a staggered potential.  The authors of Ref.\cite{Sor} proposed that a long zigzag ribbon in a staggered potential may be used as the electrodes of a tunnel junction of a spin filter.
We suggest that a gapful RGQD  may be also be used as a {\it single} electron spintronic device.  For this purpose
a small energy spacing between the gap states is desirable.  The energy spacing between the gap states
becomes smaller  for longer zigzag edges\cite{Jeong0} and for smaller values of $U$.  The energy spacing  between the gap states can be made  smaller than $ 0.01t$.   A weak  electric field will shift the energy levels of the left edge states relative to those of the right edge.   When an energy level of the left edge state
coincides with that of a right edge state an electron tunneling occurs.
By applying an  electric field it would thus be  possible to  modulate the transfer of  a spin-down  electron from one edge to the other  (on the other hand, it would be difficult to modulate the electron transfer by varying the strength of a staggered potential since it is experimentally difficult to change $\Delta$).

\section*{Acknowledgments}
This research was supported by Basic Science Research Program
through the National Research Foundation of Korea(NRF) funded by the
Ministry of Education, ICT $\&$ Future Planning(MSIP) (NRF-2015R1D1A1A01056809).


\end{document}